\documentclass[useAMS,usenatbib,referee]{biom} 

\usepackage{fancyhdr}
\usepackage[figuresright]{rotating}
\usepackage{tikz}
\usepackage{amsmath}
\usepackage[hyphens]{url} 
\usepackage{hyperref}
\usepackage[utf8]{inputenc}
\usepackage{graphicx}
\usepackage{subfig}
\usepackage{longtable}
\usepackage{booktabs}
\usepackage{multirow}
\usepackage{multicol}
\usepackage{colortbl}
\usepackage{hhline}
\usepackage{array}
\usepackage{hyperref}
\usepackage{float}
\usepackage{wrapfig}
\usepackage{lscape}
\newlength\Oldarrayrulewidth
\newlength\Oldtabcolsep

\providecommand{\tightlist}{%
  \setlength{\itemsep}{0pt}\setlength{\parskip}{0pt}}

\title[DAGs for interpretation of latent class analysis]{Improving interpretation of latent class models for
diagnostic tests by recognizing their measurands via directed acyclic graphs (DAGs)}

\author{
Nandini Dendukuri$^{1,*}$, Ian Schiller$^{2}$, Else Bijker$^{3}$,\\
\textbf{Michael Libman$^{4}$, Paul Gustafson$^{5}$, Patrick Bossuyt$^{6}$ and Joanna Merckx$^{7}$}\\ \\
1: Department of Medicine, McGill University 
2: Research Institute - McGill University Health Centre, \\
3: University of Oxford,
4: J. D. Maclean Centre for Tropical Medicine, \\
5: University of British Columbia,
6: Amsterdam Medical Centre, \\
7: Department of Epidemiology, Biostatistics and Occupational Health, McGill University}
 
\begin{document}

\date{{\it Received Nov} 2023}

\pagerange{\pageref{firstpage}--\pageref{lastpage}} \pubyear{2023}

\volume{0}
\artmonth{January}
\doi{0000-0000-0000}


\label{firstpage}


\begin{abstract}
In the absence of a perfect diagnostic test for a target condition, multiple imperfect tests may be used to arrive at a clinical diagnosis. Latent class analysis can be used to model such data with the objective of estimating test accuracy and target condition prevalence.  Such models typically assume two latent classes - target condition positive and target condition negative. However, as we will illustrate in this manuscript, this would be an oversimplification if the different tests do not share the target condition as their measurand. We show how a Directed Acyclic Graph (DAG) can be used to illustrate the relationships between the relevant variables - the observed imperfect test results, their latent measurands, the latent target condition of interest and observed covariates - revealing any conditional dependence relations. The DAG helps determine the number of latent classes, underlying the observed data, and their labels. We show how the likelihood function changes due to incorporating the measurand of each test. We study the impact on identifiability of the model. Using simulation studies we show how ignoring the measurand of an imperfect test, when it is distinct from the target condition, can lead to biased estimates of test accuracy and prevalence. We illustrate the value of the proposed approach by re-analyzing two datasets used in previously published latent class analyses of tests for pediatric tuberculosis and leptospirosis.
\end{abstract}

%
%


\maketitle

\hypertarget{introduction}{%
\section{Introduction}\label{introduction}}

Applications of latent class models (LCMs) in diagnostic test accuracy
studies have appeared in the statistics literature since the 1980s
(\citet{Walter1988}). These models have been used to estimate disease
prevalence, test sensitivity and specificity, in the absence of a
perfect reference test for the target condition of interest. Four
decades later these applications remain limited in number, partly
because of concerns regarding the interpretability and identifiability
of LCMs (\citet{VanSmeden2014a}). In this article we propose an approach
to improve the interpretability of these models by distinguishing the
measurand of each observed diagnostic test (the quantity the test is
designed to measure) from the target condition (the quantity it is being
used to measure). We show how a directed acyclic graph (DAG) is useful
for visualizing this distinction and for informing the labeling of the
underlying latent classes. We also study how the addition of latent
variables (measurands) affects the identifiability of the model.

Clinical diagnosis of a target condition typically relies on multiple
imperfect measures, each of which may target a different measurand and
provide a different perspective (\citet{Bossuyt2020}). For example, to
diagnose an infectious disease a physician may rely on visible symptoms
(e.g.~cough), measurements made on samples of bodily fluids (e.g.~blood,
sputum) or measurements recorded via imaging (e.g.~x-ray or computerized
tomography (CT) scan). However, applications of latent class models for analysis of multiple test results typically assume that all observed tests
measure the same latent target condition and that subjects fall into two
latent classes (labelled target condition positive and negative). We
refer to this approach as the 2-latent class (2LC) model. By
recognizing the different measurands involved in an application, and
their relation with the target condition, our proposed method results in
an expansion of the 2LC model to a model which could have more latent
classes. It will also serve to select an appropriate conditional
dependence structure between the tests.

In Section 2 we introduce two motivating examples based on previously
published 2LC models and show how they can be illustrated using directed acyclic graphs (DAGs). In
Section 3 we introduce the likelihood function of an LCM that accounts for the measurand of each test and review criteria for examining if the model is identifiable. In Section 4 we use simulations to illustrate conditions under which
ignoring the measurand of a test, when it is different from the target condition, can result in incorrect
labelling of the latent classes and biased estimates of the parameters of interest. In Section 5 we return to our
motivating examples and apply the proposed methods. We conclude with a
Discussion in Section 6.

\hypertarget{motivating-examples}{%
\section{Motivating examples}\label{motivating-examples}}

We will rely on the theory and notation for DAGs, which have been described primarily in the context of causal inference problems in epidemiology (\citet{Greenland1999}). We depict latent variables by ovals and observed variables by rectangles. Each observed diagnostic test result has a latent measurand, depicted by an oval with a white background. The target condition of interest, which typically causes the measurands, is depicted by an oval with a blue background. Measurands may also be caused by conditions besides the target condition of interest, depicted by ovals with a light grey background. As illustrated in the examples below, we will start off with a more complex DAG which can be simplified by removing associations which are redundant or which cannot be estimated with the available data.

\hypertarget{pediatric-pulmonary-tuberculosis}{%
\subsection{Pediatric pulmonary
tuberculosis}\label{pediatric-pulmonary-tuberculosis}}

Results of five imperfect tests were gathered from a pediatric cohort presenting with symptoms of pulmonary tuberculosis (TB) or active TB, 
(\citet{Schumacher2016})(Figure 1). The measurand for both culture and smear microscopy is the ``\emph{Mycobaterium tuberculosis (M.tb)} bacteria''. The Xpert test (T3) is designed to measure ``DNA of \emph{M.tb} bacteria''. Chest x-ray measures intrathoracic abnormalities attributable to TB or other respiratory diseases. The Tuberculin Skin Test (TST) measures an immune reaction resulting from a previous exposure to mycobacteria due to TB or another infection. The covariates age, human immunodeficiency virus (HIV) and malnutrition, which affect the risk of both TB infection and active TB were observed. 

Figure 1a presents the more detailed DAG  for the pediatric pulmonary TB data. Figure 1b is a reduced version retaining only the measurands which are distinct from the target condition and statistically identifiable with the available data. We dropped the measurand ``DNA of \emph{M.tb} bacteria'' as we considered it was not distinguishable from its parent node ``Detectable \emph{M.tb} bacteria'', as the cohort did not include subjects who had a history of TB in whom these two measurands can be distinct. Three tests (culture, Xpert and Smear) share the same measurand (Detectable \emph{M.tb} bacteria), thus illustrating they are are conditionally dependent, i.e. dependent given the target condition (\cite{Wasserman2010}). When a subject has a high (or low) bacterial load all three tests are more likely to identify them correctly as having TB (or miss them). We dropped the target conditions ``Other respiratory disease'' and ``Other infection'' as the available observed data does not provide enough information to estimate their association with the measurands caused by them, or their association with each other. Accordingly, in Figure 1b, we relabeled the measurand of x-ray as ``Intrathoracic abnormalities due to TB'' and the measurand of TST as ``Immune response to \emph{M.tb}''. 

\hypertarget{leptospirosis}{%
\subsection{Leptospirosis}\label{leptospirosis}}

In a cohort of adults suspected of leptospirosis infection, presenting in a hospital setting in Tanzania, results of four diagnostic tests were gathered (\citet{Schofield2021})(Figure 2). Of these tests, three were rapid antibody tests that picked up IgM antibodies which appear shortly after an individual is infected. The fourth test was a microscopic agglutination test (MAT) that involves testing at two points in time (at the time of recruitment and two weeks later) to pick up IgM antibodies as well as IgG antibodies, which appear at a later point in time. 

Figure 2a presents a DAG which shows that IgM antibodies may also be caused by an infection besides the target condition of Leptospirosis infection. Figure 2b presents the reduced DAG which we will fit as we do not have any observed tests of the other infection. We can see that the three rapid tests, which share the same measurand, are dependent conditional on the \emph{Leptospirosis} status due to their measurand, IgM.

\hypertarget{latent-class-models-distinguishing-the-measurands-of-the-observed-tests-and-the-target-condition-of-interest}{%
\section{Latent class models for distinguishing measurands of the observed tests from the target condition}\label{latent-class-models-distinguishing-the-measurands-of-the-observed-tests-and-the-target-condition-of-interest}}

\hypertarget{notation}{%
\subsection{Notation}\label{notation}}

Let \(T_1, \ldots T_J\) denote \(J\) tests that are used to  diagnose a target condition \(D\). We assume that each test measures one of \(P\) different measurands \(M_1, \ldots , M_P\), \((P <= J)\). To start, we will assume tests, measurands and the target condition are dichotomous with a value of 1 denoting positive and 0 denoting negative.

Assuming the tests are conditionally independent given both their measurand and the target condition, their joint probability may be expressed as follows:
\begin{align}
\label{eqn1}
P(T_1, \ldots T_J) &= P(T_1, \ldots T_J | D=1) P(D=1) + P(T1, \ldots, TJ | D=0)P(D=0) \nonumber \\
&= \sum_{M_1=0}^1 \sum_{M_2=0}^1 \ldots \sum_{M_P=0}^1 P(T_1, \ldots T_J |M_1, \ldots, M_P, D=1) P(M_1, \ldots, M_P, D=1) \nonumber \\
& \hspace{3em} + P(T_1, \ldots T_J | M_1, \ldots, M_P,D=0)P(M_1, \ldots, M_P, D=0) \nonumber \\
&= P(D=1) \sum_{M_1=0}^1 \sum_{M_2=0}^1 \ldots \sum_{M_P=0}^1 P(T_1 | M_1) \ldots P(T_J | M_P) \prod_{p=1}^P P(M_p | D=1) \nonumber \\ 
& \hspace{3em}  + P(D=0) \sum_{M_1=0}^1 \sum_{M_2=0}^1 \ldots \sum_{M_P=0}^1 P(T_1 | M_1) \ldots P(T_J | M_P) \prod_{p=1}^P P(M_p | D=0)  
\end{align}.

The terms \(P(T_j|M_p)\) provide the accuracy of the
\(j^{th}\) test with respect to the \(p^{th}\) measurand and the terms \(P(M_p|D)\) provide the accuracy of the \(p^{th}\) measurand with respect to the target condition. The prevalence of the target condition is given by
\(P(D=1)\).

The prevalence of a measurand \(M_j\) can be calculated as
\(P(M_p = 1) = P(M_p = 1|D=1)P(D=1) + P(M_p = 1 |D=0)P(D=0).\) The sensitivity of the
\(j^{th}\) test with respect to the target condition can be calculated as
\(P(T_j=1| D=1) = P(T_j |M_p=1)P(M_p=1|D=1) + P(T_j|M_p=0) P(M_p=0|D=1)\). Similarly, the specificity would be \(P(T_j=0| D=0) = P(T_j |M_p=1)P(M_p=1|D=0) + P(T_j|M_p=0) P(M_p=0|D=0)\)

\hypertarget{conditional-dependence}{%
\subsubsection{Conditional dependence}\label{conditional-dependence}}
When multiple diagnostic tests share the same measurand in (\ref{eqn1}) they are dependent conditional on the target condition, $D$. Multiple measurands may also be conditionally dependent if their accuracy is a function of another variable, $r$ conditional on the target condition. The variable that causes the conditional dependence (measurand or $r$)  may be dichotomous, ordinal or continuous (\cite{Wang2016}).  

We can express the measurand as a random effect \(M_{pd}\) such that \(P(T_j | M_{pd}, D=d) = g_M(a_{jpd} + b_{jpd} M_{pd}), M_{pd} \sim f_M(.)\) (\cite{Collins2014,Dendukuri2001}), where the additional index \(d\) is added to recognize that the random effect \(M_{pd}\) may be dependent on the target condition status and where \(g_M\) is a suitable link function, e.g.~ the cumulative logistic or normal probability density function, and $f_M(.)$ is a suitable probability density function for the measurand. Likewise, we could introduce a random effect (\(r_d\)) with a probability density function \(f_r(.)\) and express the accuracy of each measurand as \(P(M_{pd} | D=1, r_d) = g_r(u_{pd} + v_{pd} r_d), r_d \sim f_r(.)\), where \(g_r\) is a suitable link function.  

\hypertarget{identifiability}{%
\subsection{Identifiability}\label{identifiability}}

Latent class models are susceptible to non-identifiability, i.e.~the
existence of multiple solutions for the unknown parameters
(\citet{Walter1988}). Two necessary conditions for identifiability are: 1) the number of degrees of freedom exceeds
the number of unknown parameters, 2) the Jacobian of the
transformation of the multinomial probability vector for the observed
data to the vector of parameters being estimated is non-zero (a measure
of local, though not global, identifiability) (\citet{Goodman1974}).

The number of degrees of freedom available from observing \(J\)
dichotomous tests is \(2^J – 1\). Assume there are \(P\) 
measurands included in the latent class model which are distinct from the target condition. The number of unknown parameters to be
estimated, when assuming the only variables involved are the tests,
measurands and target condition, is \(1 + 2P + 2J\) (prevalence of the
target condition + 2 accuracy parameters for each measurand with respect
to the target condition + 2 accuracy parameters for each test with
respect to its measurand). If random effects or covariates are involved, the number of unknown parameters may increase further. 

\hypertarget{bayesian-inference}{%
\subsection{Bayesian inference}\label{bayesian-inference}}

The latent class models described above can be estimated using either
classical or Bayesian inference. Bayesian inference will be used in this
manuscript as its utility encompasses the situation where prior information is necessary to obtain a solution in the event the model is not identifiable
(\citet{Dendukuri2001},\citet{Gustafson2005}). Bayesian inference
requires that a prior distribution be provided for each unknown
parameter. For parameters such as prevalence or sensitivity/specificity,
which lie between 0 and 1, a \(Beta(\alpha, \beta)\) prior distribution is a suitable choice. Values of \(\alpha\) and \(\beta\) can be chosen so the distribution is non-informative (e.g.~\(\alpha=\beta=1\)) or
informative. When including a random effect in the model, Normal(0, 1) prior distributions would be suitable for the intercept parameters (\(a_{jpd}\) and \(u_{pd}\)) and Half-Normal(0,1) prior distributions for the slope parameters (\(b_{jpd}\) and \(v_{pd}\)). Section 5 provides
specific examples. Bayesian inference is carried out using Markov Chain Monte Carlo (MCMC) algorithms. We will
implement MCMC methods using the rjags package (\citet{rjags2022}) in the R software environment (\citet{R2022}). 

Once we are satisfied that the MCMC process has converged, we check if the model fits the observed data well by comparing the observed and expected frequencies for each pattern of test results. We examine the correlation residuals (observed-expected pairwise correlation) (\citet{Qu1996}) to see if there is evidence of conditional dependence that has not been adjusted for. Once we are satisfied the model fits the observed data, we report on the posterior distribution of the parameters of interest using standard approaches for reporting Bayesian inference such as density plots and posterior quantiles.

\hypertarget{simulation}{%
\section{Simulation}\label{simulation}}

In this section we carry out a simulation study to demonstrate the
impact of ignoring the measurands of the observed tests when they are
distinct from the target condition. 

\hypertarget{simulation-settings}{%
\subsection{Simulation settings}\label{simulation-settings}}

We simulated data based on a series of DAGs for the results of five dichotomous tests. We assumed that for some of the tests their measurand was the target condition (TC) itself, while other tests had a measurand (M) distinct from the target condition. 
The number of tests which measure TC was allowed to range from one to four. Figure 3a below illustrates the DAG for the case when four tests measure TC and only one measures M. Conversely, in Figure 3b, only one test measures TC and four tests measure M.

For each DAG the following simulation settings were used, inspired by values plausible in diagnostic accuracy studies of infectious diseases. All tests for TC had sensitivity of 80\% and
specificity of 98\%. All tests for M had a sensitivity of 70\% and specificity of 98\%, with respect to M. The prevalence of TC was fixed at 15\%. For each DAG, we generated different scenarios by varying the sensitivity of
M with respect to TC from 0\% to 100\%, while keeping the specificity of M with respect to TC fixed at 87.6\%. This resulted in the prevalence of M varying from 10.5\% to 25.5\% across the scenarios. In each scenario we used the expected dataset (resulting from
multiplying the sample size by the probability of each test pattern) to eliminate the effect of random sampling variation.

The latent class models corresponding to all simulated DAGs have four latent classes resulting from possible combinations of TC and M (TC+M+, TC+M-, TC-M+, TC-M-). We fit the 2LC model to each simulated dataset to study the impact of ignoring M and labeling the latent classes as (TC+, TC-). We then compared the estimates, of the latent class prevalence and accuracy of each test with respect to TC, to the true values.

\hypertarget{simulation-results}{%
\subsection{Simulation results}\label{simulation-results}}

We describe here the results of the simulation exercise corresponding to
the two DAGs in Figures 3a and 3b. Similar results for the remaining DAGs, appear in the supplementary material.

Datasets generated using the DAG in Figure 3a resulted in a TC prevalence estimate of around 15\% based on the 2LC model, close to the true prevalence of TC. Further, the sensitivity of tests 1-4 was close to the true value of 80\% (see Figure 4a). Thus the 2LC model appears to correctly identify latent classes which are TC+ and TC- in this situation. Figure 5a also shows that the estimated mean sensitivity of the fifth test decreased steadily as the sensitivity of M decreased. When the sensitivity of M was 100\%, i.e.~a situation where TC and M are
virtually the same, the fifth test's sensitivity was estimated close to
its true value (with respect to M) of 70\%. However, as we
deviate from this case the estimate that is provided by the model for
the fifth test is in fact the \textbf{mean} sensitivity estimate with respect to TC.

On the other hand, for data generated using the second DAG in Figure 3b,
where the majority of tests were designed to measure M, the prevalence estimate from the 2LC model was close to the
true prevalence of M (see supplementary material). We can
see in Figure 4b that the sensitivity estimates for tests 2-5 based on
the 2LC model are roughly 70\%, which is in fact their sensitivity with respect to M. The estimated sensitivity of the first test lies well below the true value of 80\% for all the scenarios generated. This is because, the latent classes identified by the 2LC model are M+ and M-. Therefore the estimate of the sensitivity of \(T_1\) reported by the model corresponds to its sensitivity with respect to M.

Other simulated scenarios were in between these two extremes. This exercise illustrates that even though all the tests directly or indirectly measure TC, the 2LC model will not always
return the desired classification of target condition positive and
negative. Rather the resulting estimates will
depend on the structure of the data generating model and whether the measurands of the observed tests differ from the target condition. Of course, fitting a model with 4 latent classes with constraints reflecting whether tests measure TC or M, returns unbiased estimates of all parameters (results not shown).

\hypertarget{applications}{%
\section{Applications}\label{applications}}

In this section we return to the motivating examples introduced earlier.
In each case, we contrast the results based on a 2LC model with those
obtained from a model with a larger number of classes based on the DAGs
in Figures 1b and 2b. The model, prior distributions used, convergence
diagnostics, goodness-of-fit diagnostics and density plots for all
models are given in the supplementary material. Here we report the
posterior median and 95\% credible interval for each parameter of
interest based on the different models.

\hypertarget{pediatric-pulmonary-tuberculosis-1}{%
\subsection{Pediatric pulmonary
tuberculosis}\label{pediatric-pulmonary-tuberculosis-1}}

\textbf{Determining the number of latent classes:} Table 1 shows how combinations of target condition and the two dichotomous measurands in Figure 1b - active TB, latent TB and intrathoracic
abnormalities due to TB - result in four latent classes. The measurand ``Detectable TB Bacteria" is treated as continuous and therefore does not appear as a column in the table.  Since we are
assuming that latent TB causes active TB, the combinations where a child has active TB but not latent TB are not possible (NP). Likewise, since we are assuming active TB causes intrathoracic abnormalities due to TB, classes where the latter is present but not the former are not possible.

Table 1 also lists the probability of a positive result on each observed test within each latent class expressed in terms of the accuracy parameters. For example, in latent class 1 the target condition and both measurands are positive, and the probability of a positive test for each observed test is its sensitivity with respect to its measurand. In the case of Culture, Xpert and Smear, accuracy parameters are with respect to the target condition, aggregated over the distribtution of their continuous measurand. 

\textbf{Identifiability:} We can deduce from Table 1 that there are 13 unknown parameters to
be estimated including the prevalence of the different latent classes (3
parameters) and the test accuracy (10 parameters). The number of
available degrees of freedom is 31. As in the original publication
(\citet{Schumacher2016}), we allowed prevalence of the latent classes
to be functions of the child's HIV and malnutrition status. Taking into account 1 additional parameter due to the presence of a continuous measurand, the total number of unknown parameters was 19, implying the model satisfied the first criterion that the number of parameters is less than the number of degrees of freedom available. The second criterion of local identifiability was not satisfied. The rank of the Jacobian was 12 whereas the number of unknown parameters was 14. Therefore at least 2 informative prior distributions are needed to obtain a meaningful solution.

\textbf{Prior distributions:} As in the earlier publication by \cite{Schumacher2016}, based on expert input prior information was provided on the specificity of
culture \((0.99 \leq S_C \leq 1)\) and the specificity of Xpert
\((0.98 \leq S_C \leq 1)\), while non-informative prior distributions were used for the remaining parameters. The resulting model did not exhibit any problems with convergence and the expected values of the frequency of test patterns and pairwise correlation did not differ significantly from the observed values (supplementary material).

\textbf{Results:} The prevalence of each of the four latent classes is given in Table 1. The prevalence of active
TB (95\% credible interval (CrI)) was 0.22 (0.18,0.28)), the prevalence of latent TB was 0.49 (0.38,0.64) and the prevalence of intrathoracic abnormalities due to TB was 0.13 (0.09,0.19). The estimate of the prevalence of active TB was similar to but lower than the previously published estimate of 0.27 (95\% credible interval (CrI) 0.21, 0.35)
from the 2LC model (\cite{Schumacher2016}). This was because the
probability of having Active TB decreased for some subjects while their probability of having TB Infection increased. 
Table 2 shows that the estimated accuracy of TST and chest x-ray are higher with respect to their measurands than with respect to Active TB. Table 1A (supplementary material) provides the probability of each latent class conditional on the
test results. Positive results on either or both of x-ray and TST were predictive of
active TB only when accompanied by a positive result on at least one of
culture, Xpert or smear. Subjects who were positive on TST alone were
highly likely to be in latent class 3, i.e.~they had a high probability
of having latent TB infection. Whereas under the 2LC model, this group
had a 52\% chance of having Active TB, under the 4LC model there was
more certainty they did not have Active TB. Those who were positive on
the x-ray alone were most likely to be in latent class 4, i.e.~it is
unlikely they had any of the three measurands of interest identified in
Figure 1b.

\textbf{Face validity:} Due to the lack of a perfect reference test, we can only use indirect
information to validate this model. Data on whether each child received
TB treatment (see Table 1 extreme right column), suggests that children
classified by the latent class model as having Active TB were the most
likely to be treated, followed by children with TB Infection. Children
without Active TB or any of the measurands were least likely to be treated.

\hypertarget{leptospirosis-1}{%
\subsection{Leptospirosis}\label{leptospirosis-1}}

\textbf{Determining the number of latent classes:} Table 3 summarizes all eight possible combinations of the measurands and target condition in Figure 2. Some of the latent classes are not possible due to the prior distribution as explained below.

\textbf{Identifiability:} The number
of available degrees of freedom from four observed tests is 15. The number of parameters of interest is at least 1 + 2P + 2J
= 1 + 2(2) + 2(4) = 13, if we allowed for all eight possible latent classes. The first condition for
identifiability is satisfied with the number of unknown parameters
being less than the number of degrees of freedom. However, the second condition of local identifiability is not satisfied as the rank of the Jacobian is 12 (see supplementary material). 


The 2LC model for this problem, which ignores the measurands, has 1 + 2J = 1
+ 2(4) = 9 unknown parameters and meets both identifiability criteria. We could also simplify the
model in Figure 2b and Table 3 by ignoring the target condition D. This
would reduce the number of latent classes to four and the number of unknown parameters to 1 + P + 2J = 1 + 2 +
2(4) = 11. Though the 4LC model is also identifiable (see supplementary
material), it does not provide the accuracy of the tests with respect to
the target condition or the prevalence of the target condition.

\textbf{Prior distributions:} In order to estimate the complete model with eight latent classes, we would need to provide an informative distribution on at least one parameter as the rank of the Jacobian is one less than the number of degrees of freedom. Based on expert opinion from co-authors familiar with tests for Leptospirosis, we
used the following prior information on specificities of the two
measurands:

\begin{itemize}
\tightlist
\item
  specificity of IgM exceeds 90\%
\item
  specificity of IgG is 100\%
\end{itemize}

This amounts to removing lines 5 and 7 of Table 3 and reducing the
number of latent classes to six. The assumption behind the previously fit
2LC model (\citet{Schofield2021}) is that all four tests are directly
measuring Leptospirosis, while the 6LC model recognizes that the three
rapid tests measure IgM only, while the MAT test measures both IgG and
IgM. The 6LC model separately estimates each test's accuracy with
respect to its measurands, as well as the accuracy of each measurand
with respect to the target condition.

\textbf{Results:} The prevalence of each latent class is given in
Table 3. The prevalence of the target condition and measurands under different models are compared in Table 4. The summary test accuracy statistics are compared
in Table 5. 

From Table 4, we can see that the point estimate of the prevalence of
Leptospirosis based on the 6LC model is higher than the estimate from
the 2LC model. The estimates of the prevalence of IgM and IgG were not very precise making it unclear which of them had a greater prevalence. As
expected, the composite of IgM and IgG has a higher sensitivity with
respect to Leptospirosis than either IgM or IgG alone. All measurands
have a high specificity as expected. The wide credible intervals show
that little was learned about the sensitivity of the measurands
themselves. Further, the estimate of the prevalence of Leptospirosis was
also imprecise. The imprecision can be reduced if it is reasonable to
use more informative (i.e.~more precise) prior information.

Table 5 provides estimates of the accuracy of the tests. As reported
previously by \citet{Schofield2021} the sensitivity of the MAT test
based on the 2LC model is lower compared to the other tests. The
expansion of the model to include 6 latent classes helps to see that all
4 tests have a low sensitivity with respect to Leptospirosis.

Also, all four tests have a better accuracy with respect to their
measurand, which they were designed to measure, than with respect to
Leptospirosis. This is especially apparent for both Leptocheck and
Test-IT, which measure IgM.

The prevalence of IgM based on the 6LC model is similar to the
prevalence of Leptospirosis based on the 2LC model (Table 4). Further,
the sensitivity and specificity estimates from the 2LC model bear a
striking resemblance with the sensitivity and specificity estimates with
respect to the their measurands from the 6LC model (Table 5). We can
conclude that the latent class identified by the 2LC model appears to be
IgM, which was the measurand for the majority of tests, and not
Leptospirosis infection.

\hypertarget{discussion}{%
\section{Discussion}\label{discussion}}

Latent class models (LCMs), or finite mixture models, have been widely
applied in clinical research. In some cases, they are used as
exploratory tools to identify latent clusters. However, in the case of
diagnostic test accuracy research, their role is confirmatory as the
latent classes of interest are known ahead of time to be target condition positive and negative. Whereas most applications of LCMs in diagnostic accuracy
studies use a model that assumes only these two latent classes, our manuscript shows
how this approach ignores the latent measurand of each observed diagnostic test and the
possibility that there are more than two latent classes across which we
can distinguish the changes in each test's accuracy.

A major criticism of LCMs for diagnostic accuracy estimation is that
LCMs with different dependence structures may be indistinguishable using
traditional measures of model fit, but can result in different accuracy
estimates. This implies one could easily make the mistake of using the wrong
model.  Several authors have recognized the
difficulty in selecting the ``correct'' latent class model for a given
application (\citet{Schofield2021},\citet{Albert2004}). 
To address this problem, we have proposed the usage of a directed
acyclic graph (DAG), a tool that is increasingly accepted in the
clinical research literature on causal inference. In the context of
diagnostic accuracy research, a DAG serves to illustrate the additional
variables, besides the target condition, that are involved. It also
aids in transparent communication between clinicians, epidemiologists
and biostatisticians for defining the latent class model and reporting
it. Rather than fitting multiple models and selecting among them, we advocate fitting a single model that best reflects our knowledge of the problem. Nonetheless model selection approaches may be relevant for evaluating aspects of the model that we do not have any information about.

Further, as we have demonstrated in our applied examples, it is
important to go beyond the DAG to illustrate how the combination of
latent variables results in latent classes that can be identified. This
also helps to label the latent classes appropriately. As illustrated in
our simulations, the classes identified by the traditional 2LC model are
determined by the combination of observed tests. If the majority of
tests do not measure the target condition of interest, it is possible
that the latent classes are mislabeled in an 2LC model.

The applied examples illustrate that we can distinguish between a test's
accuracy with respect to its' measurand and with respect to the target
condition. When the two are not the same, the sensitivity and
specificity with respect to the target condition is affected by the
prevalence of the measurand. For example, in the Leptospirosis example,
the low prevalence of IgG and IgM resulted in a low sensitivity of the
tests to detect the target condition. However, the same tests had higher
accuracy with respect to their measurand, which is what they were
designed to measure. This observation also provides face validity for
the latent class structure we have employed.

In an earlier analysis of the Leptospirosis data, \citet{Schofield2021}
commented on the surprisingly low estimate of sensitivity of the MAT
test based on the 2LC model. From the 6LC model we reported, we can see
that in fact the measurand of MAT had a higher sensitivity (than the
measurand of the rapid tests) as it detects both IgG and IgM. But this
should not be confused with the sensitivity of the MAT test itself with respect to the target condition, which is a function of the prevalence of this measurand. The low prevalence in this sample resulted in a low sensitivity of all four tests.

As shown in our simulations, when a majority of tests share the same
measurand the 2LC model will identify that measurand. This explains why
the results of the 2LC model for the TB example identified the class of
active TB as three of the five tests shared this measurand. On the other
hand, in the Leptospirosis example the 2LC model identified the class of
IgM positive, which was the target for three out of four of the tests.

The proposed approach is not without limitations. We assumed that the
observed tests are dichotomous, whereas in practice they can be
continuous. Future work should consider this possibility and examine
whether it further improves estimates of diagnostic accuracy and
prevalence. Another limitation is that we did not model missing data in
either example. This should be possible by extending the model to allow for missing values to be imputed \cite{MacLean2023}. 

\hypertarget{figures-and-tables}{%
\section{Figures and tables}\label{figures-and-tables}}

\hypertarget{figures}{%
\subsection{Figures}\label{figures}}

\begin{figure}
 \centering 
  \subfloat[Complete version]    {\includegraphics[width=0.75\linewidth,keepaspectratio]{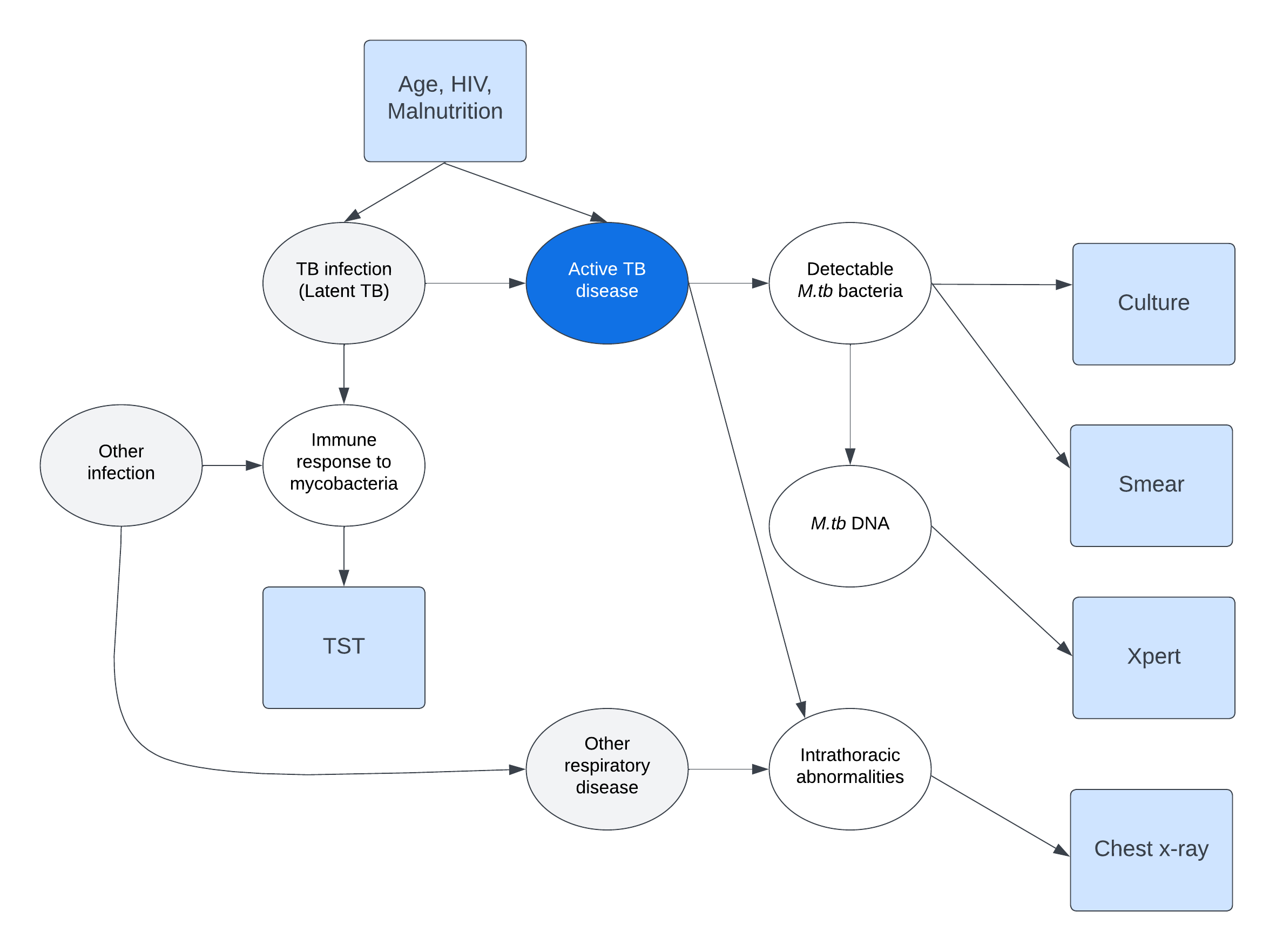} 
    \label{fig:fig1a}}
\hfill
  \subfloat[Reduced version]{
    \includegraphics[width=0.5\linewidth,keepaspectratio]{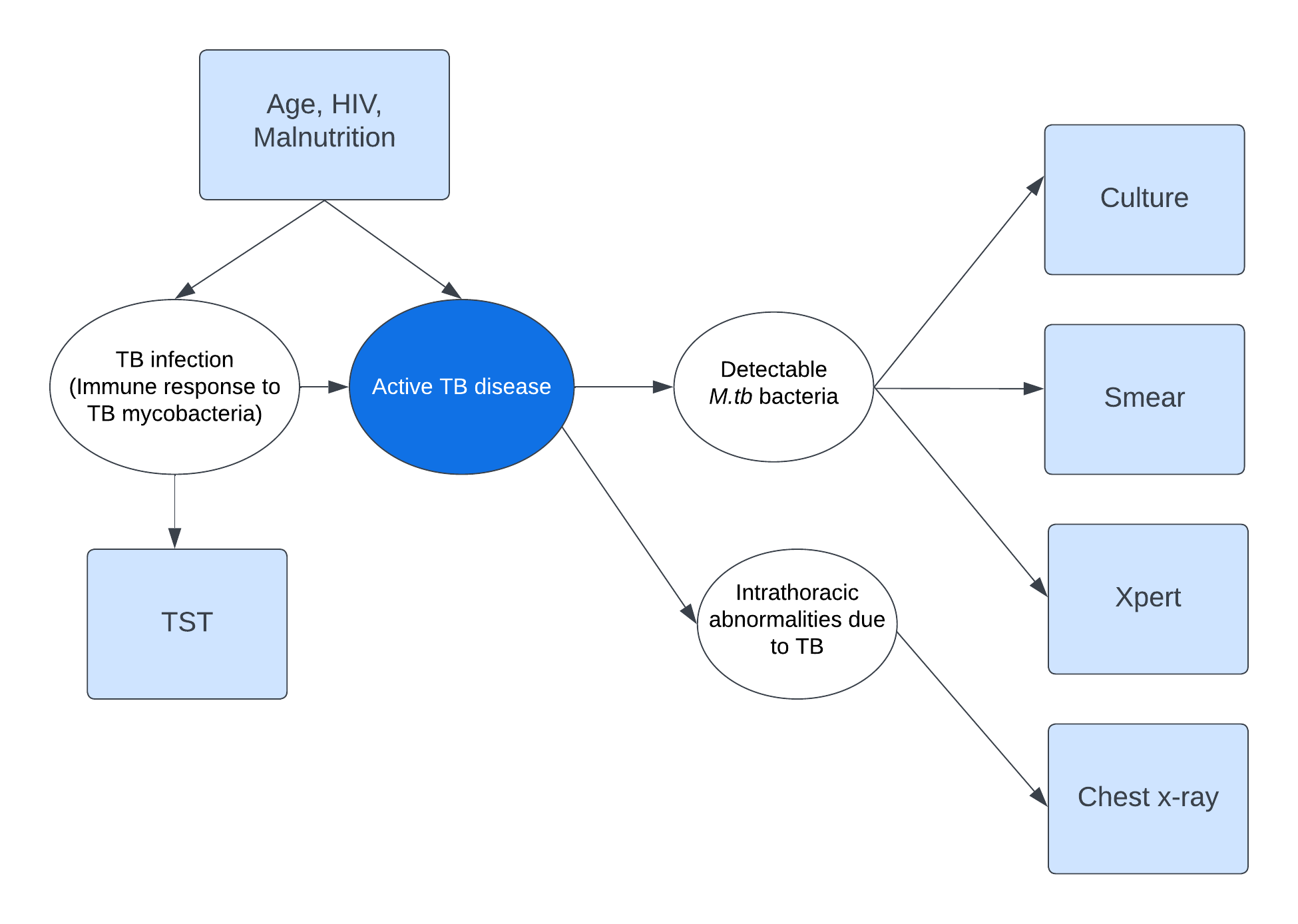} 
    \label{fig:fig1b}}
    \caption{Directed Acyclic Graph (DAG) for pediatric pulmonary tuberculosis tests}
\label{fig1}
\footnotesize{Active TB disease refers to pediatic pulmonary tuberculosis}
\end{figure}

\begin{figure}
\centering 
\subfloat[Complete]{\includegraphics[width=0.5\linewidth]{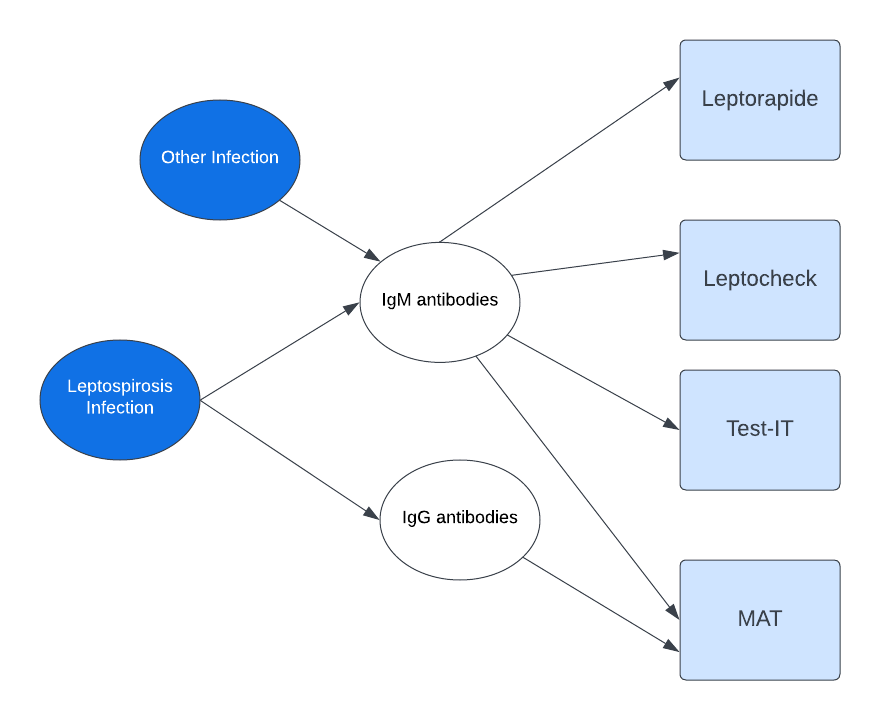}\label{fig:fig2a}}
\hfill
\subfloat[Reduced]{\includegraphics[width=0.5\linewidth]{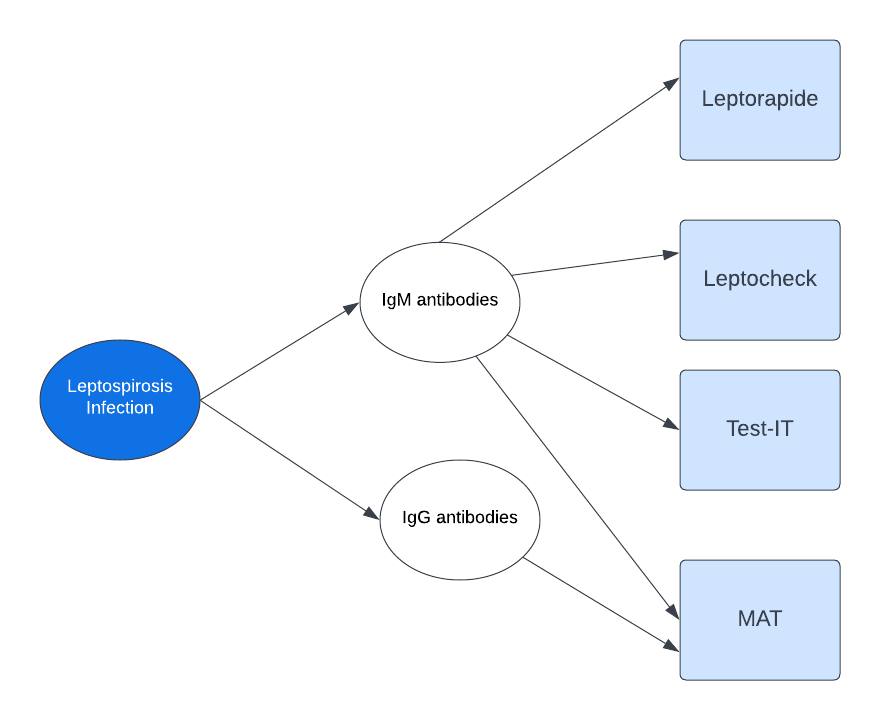}\label{fig:figb2}} 
\caption{Directed Acyclic Graph for Leptospirosis tests}
\label{fig2}
\end{figure}

\begin{figure}
\centering 
\subfloat[4 tests that measured the target condition vs 1 test that measured the non-specific measurand]{\includegraphics[width=0.5\linewidth]{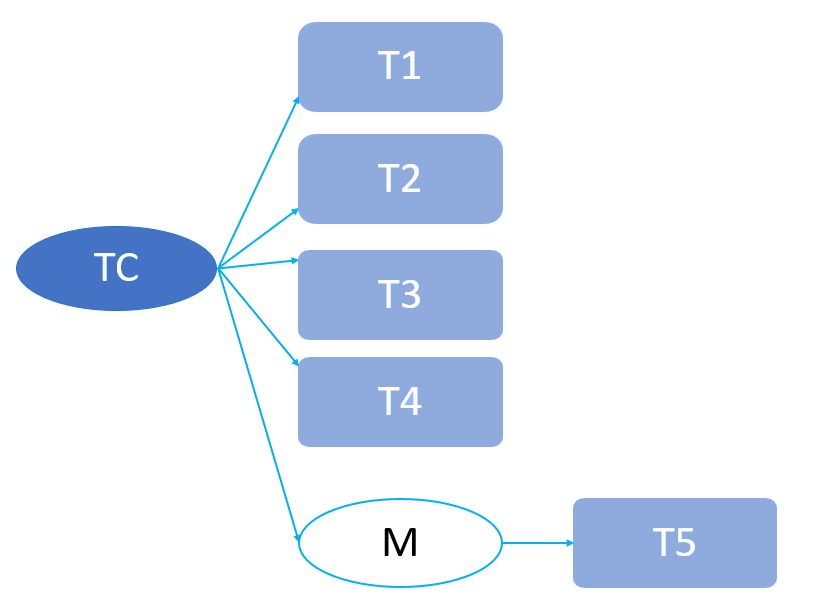}\label{fig:fig3a}} 
\hfill
\subfloat[1 tests that measured the target condition vs 1 test that measured the non-specific measurand]{\includegraphics[width=0.5\linewidth]{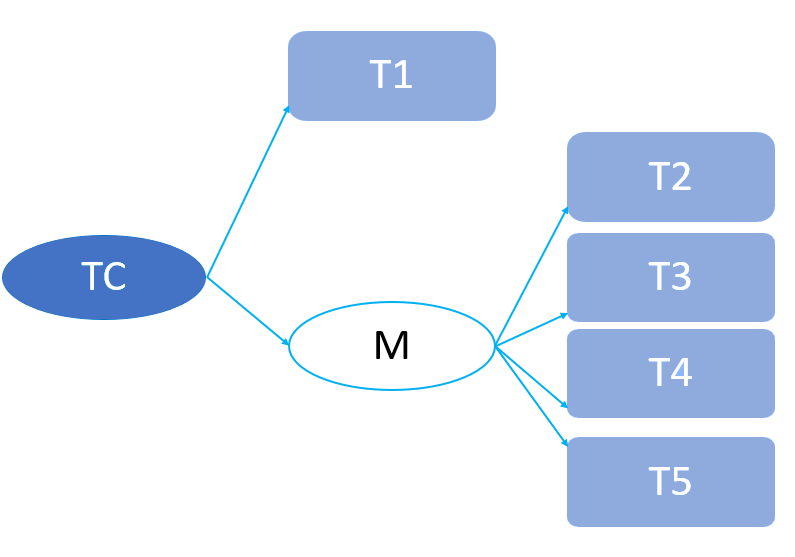}\label{fig:fig3b}} 
\label{fig3}
\caption{Directed Acyclic Graph for simulation}
\end{figure}

\begin{figure}
\centering 
\subfloat[Majority of tests measure target condition]{\includegraphics[width=0.5\linewidth]{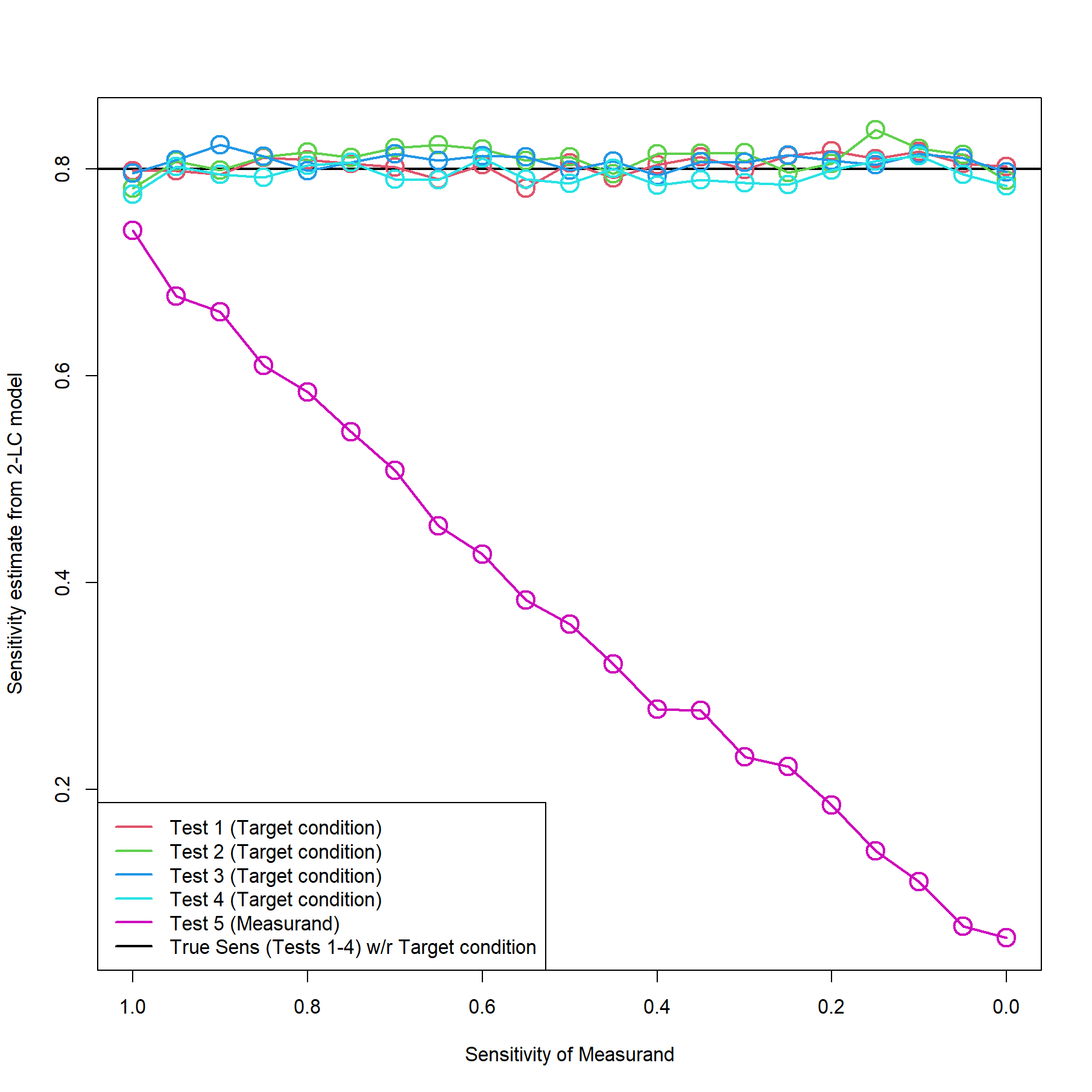}} \hfill
\subfloat[Majority of tests measure non-specific measurand]{\includegraphics[width=0.5\linewidth]{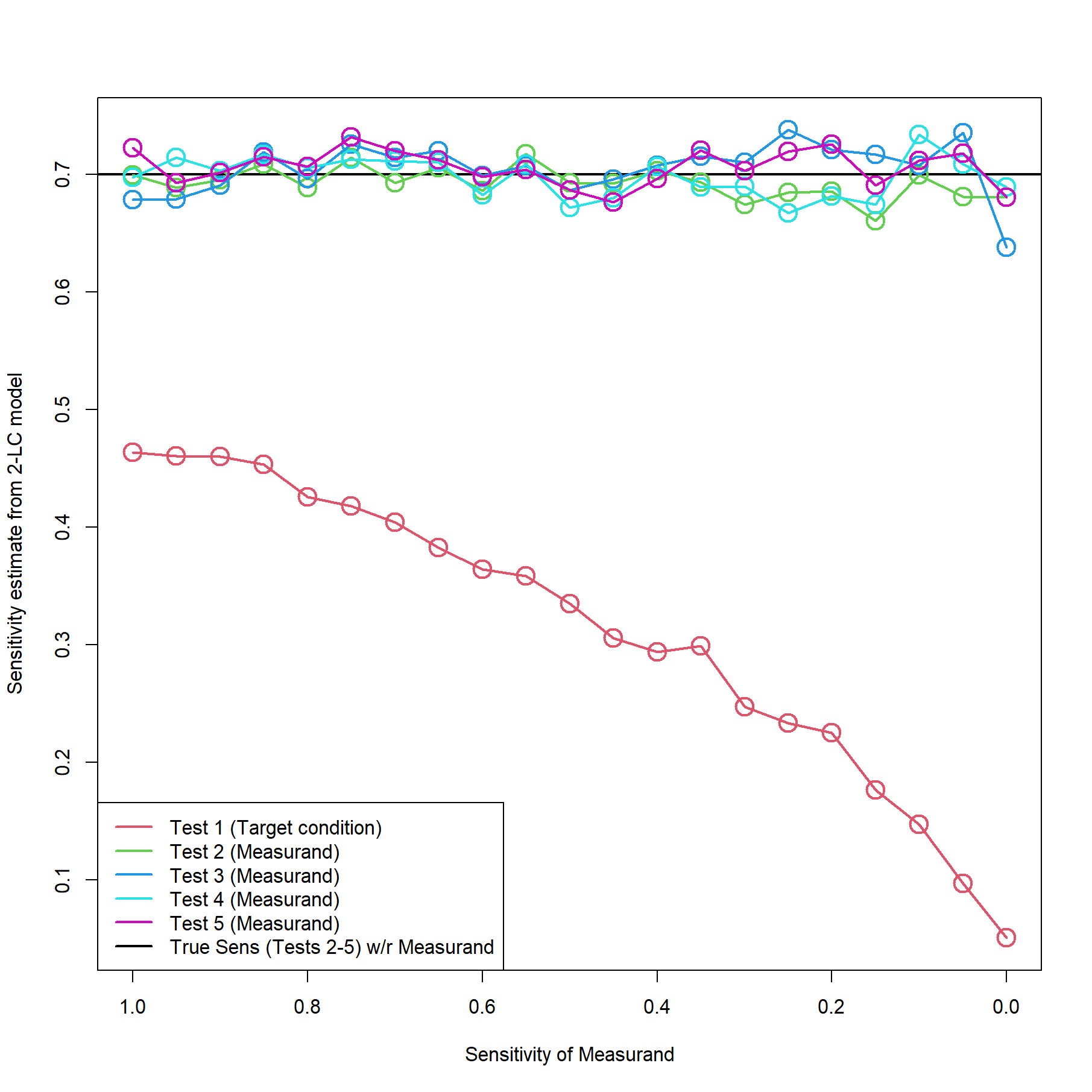}\label{fig:fig4b}} 
\caption{Sensitivity estimates from a 2-latent class model}
\footnotesize{These graphs were generated using a random dataset and not the expected dataset in order to differentiate between the estimates of the sensitivity parameters which were identical}
\label{fig4}
\end{figure}

\clearpage

\hypertarget{tables}{%
\subsection{Tables}\label{tables}}

\global\setlength{\Oldarrayrulewidth}{\arrayrulewidth}
\setlength{\tabcolsep}{2pt}
\renewcommand*{\arraystretch}{0.75}

\begin{landscape}
\begin{longtable}{ccccccccccc}
    \caption{Relation between measurands, latent classes and test accuracy parameters for TB example} \label{tab1} \\
    \hline
        Target & \multicolumn{2}{c}{Measurand} &  Latent & \multicolumn{5}{c}{Probability of positive test} & \multicolumn{2}{c}{Results} \\ \cline{2-3} \cline{5-9} \cline{10-11}
        Condition  &	TB Infection &	Intrathoracic &	Class & Culture & Xpert & Smear & Xray & TST &	Prevalence & 	Probability of  \\
Active TB & &  Abnormalities & & & & & & &  & treatment \\
& &  due to TB &  & & & & & & &  
 \\\hline
1	&	1	&	1	&	1	&	S1	&	S2	&	S3	&	S4	&	S5	&	0.13 (0.09,0.19)	&	0.94	\\
1	&	1	&	0	&	2	&	S1	&	S2	&	S3	&	1-C4	&	S5	&	0.1 (0.05,0.14)	&	0.91	\\
1	&	0	&	1	&	NP	&		&		&		&		&		&		&		\\
1	&	0	&	0	&	NP	&		&		&		&		&		&		&		\\
0	&	1	&	1	&	NP	&		&		&		&		&		&		&		\\
0	&	1	&	0	&	3	&	1-C1	&	1-C2	&	1-C3	&	1-C4	&	S5	&	0.25 (0.13,0.42)	&	0.66	\\
0	&	0	&	1	&	NP	&		&		&		&		&		&		&		\\
0	&	0	&	0	&	4	&	1-C1	&	1-C2	&	1-C3	&	1-C4	&	1-C5	&	0.52 (0.34,0.64)	&	0.4	\\ \hline
\end{longtable}
\end{landscape}

\begin{landscape}
\begin{longtable}{ccccccc}
    \caption{Estimates of TB prevalence and TB diagnostic test accuracy} \label{tab2} \\    \hline
Test &  TB Infection &	Active TB &	Intrathoracic &   Active TB & TB infection &	Intrathoracic \\ 
 &   &	 &	Abnormalities &    &  &	Abnormalities \\ \hline
& \multicolumn{3}{c}{Sensitivity} & \multicolumn{3}{c}{Specificity} \\ \hline
Culture	& 0.39 (0.29,0.51) & 0.69 (0.55,0.84) 	& 0.69 (0.55,0.84) 	&	1 (0.99,1) &	1 (0.99,1)	     &   0.92 (0.89,0.95)	 \\
Xpert	& 0.32 (0.24,0.42) & 0.57 (0.44,0.7)	& 0.57 (0.44,0.69)	&   0.99 (0.97,0.99) &	0.99 (0.97,0.99) &	0.93 (0.9,0.95)	  \\
Smear	& 0.15 (0.1,0.2)   & 0.26 (0.18,0.35)	& 0.26 (0.19,0.34)	&	1 (0.99,1) &	1 (0.99,1)	     &   0.97 (0.96,0.98) \\
X-ray	& 0.48 (0.41,0.57) & 0.66 (0.57,0.73)	& 0.95 (0.82,1)	    &	0.76 (0.72,0.81) &	0.76 (0.72,0.81) &	0.76 (0.72,0.81) \\
TST	    &   0.7 (0.56,0.82) & 0.7 (0.56,0.82)	& 0.7 (0.56,0.82)	&	0.83 (0.74,0.91) &	0.65 (0.6,0.71)	 &   0.62 (0.58,0.66) \\ \hline
\end{longtable}
\end{landscape}

\begin{landscape}
\begin{longtable}{ccccccccc}
    \caption{Relation between measurands, latent classes and test accuracy parameters for Leptospirosis example} \label{tab3 } \\
    \hline
        \multicolumn{3}{c}{Measurand} &  Latent & \multicolumn{4}{c}{Probability of positive test} & Results \\ \cline{1-3} \cline{5-9}
        Leptospirosis &        IgM &    IgG &     Class & MAT & Leptorapide & Leptochek & Test-IT & Prevalence  \\ \hline
1              &             1              &             1              &             1              &             S1           &             S2           &             S3           &             S4                &             0.03 (0,0.12)   \\
1              &             1              &             0              &             2              &             S1           &             S2           &             S3           &             S4                &             0.04 (0,0.14)   \\
1              &             0              &             1              &             3              &             S1           &             1-C2       &             1-C3       &             1-C4                &             0.06 (0,0.37)   \\
1              &             0              &             0              &             4              &             1-C1       &             1-C2       &             1-C3       &             1-C4                &             0.08 (0,0.66)   \\
0              &             1              &             1              &             NP          &                             &                             &                             &                             &                             \\
0              &             1              &             0              &             6              &             S1           &             S2           &             S3           &             S4                &             0.04 (0,0.15)   \\
0              &             0              &             1              &             NP          &                             &                             &                             &                             &                             \\
0              &             0              &             0              &             8              &             1-C1       &             1-C2       &             1-C3       &             1-C4                &             0.67 (0.06,0.88)      \\ \hline
\end{longtable}
\end{landscape}

\begin{landscape}
    \begin{longtable}{ccccccc}
    \caption{Prevalance of target condition (Leptospirosis) and measurands (IgG, IgM, IgG or IgM), and accuracy of the measurands with respect to target condition} \label{tab4} \\
      \hline
        & \multicolumn{2}{c}{Prevalence} & \multicolumn{2}{c}{Sensitivity} & \multicolumn{2}{c}{Specificity} \\
         Measurand & 2LC & 6LC & 2LC & 6LC& 2LC & 6LC \\
         & model & model & model & model & model & model \\ \hline
          Leptospirosis           &             0.14 (0.08,0.24) &             0.28 (0.02,0.93) &                             &                             &                             &                             \\
IgM        &                             &             0.13 (0.09,0.22) &                             &             0.27 (0.03,0.94) &                             &             0.93 (0.77,1) \\
IgG         &                             &             0.10 (0,0.44)       &                             &             0.42 (0.02,0.96) &                             &             1 (1,1)       \\
IgG or IgM           &                             &             0.2 (0.11,0.51)      &                             &             0.65 (0.16,0.99) &                             &                0.93 (0.77,1)       \\
\hline
       
    \end{longtable}
\end{landscape}
 
\global\setlength{\Oldarrayrulewidth}{\arrayrulewidth}
\setlength{\tabcolsep}{2.0pt}
\renewcommand*{\arraystretch}{0.75}

\begin{landscape}
\begin{longtable}{ccccccccc}
    \caption{Accuracy of observed tests with respect to (wrt) target condition (Leptospirosis) and measurands (IgM, IgG)} \label{tab5} \\
    \hline
     & \multicolumn{2}{c}{Sensitivity wrt} & \multicolumn{2}{c}{Specificity wrt} & \multicolumn{2}{c}{Sensitivity wrt} & \multicolumn{2}{c}{Specificity wrt} \\
     & \multicolumn{2}{c}{measurand} & \multicolumn{2}{c}{measurand} & \multicolumn{2}{c}{Leptospirosis} & \multicolumn{2}{c}{Leptospirosis} \\
     \hline
Test & 2LC & 6LC & 2LC & 6LC & 2LC & 6LC & 2LC & 6LC \\ \hline
MAT       &                             &             0.26 (0.12,0.45) &                             &             0.94 (0.88,0.99) &             0.27 (0.13,0.47) &                0.17 (0.09,0.37) &             0.92 (0.87,0.95) &             0.92 (0.86,0.98) \\
Leptorapide        &                             &             0.35 (0.18,0.57) &                             &             0.85 (0.8,0.9)      &             0.34 (0.18,0.56)          &             0.20 (0.13,0.39)   &             0.85 (0.8,0.9)      &             0.84 (0.78,0.89) \\
Leptochek           &                             &             0.93 (0.69,1)       &                             &             0.9 (0.84,0.98)   &             0.93 (0.69,1)                &             0.32 (0.13,0.87) &             0.9 (0.84,0.98)   &             0.84 (0.72,0.94) \\
Test-IT   &                             &             0.78 (0.47,0.99) &                             &             0.95 (0.91,0.99) &             0.77 (0.46,0.99) &                0.24 (0.07,0.76) &             0.95 (0.91,0.99) &             0.9 (0.79,0.96)   \\ \hline    
   \end{longtable}   
\end{landscape}

\newpage
\bibliographystyle{biom}
\bibliography{references.bib}

\label{lastpage}

\end{document}